\documentclass[aps,prb,showpacs,groupedaddress,amssymb,twocolumn]{revtex4}
\usepackage{mathptmx}
\usepackage{bm,amsmath}  %%% do not include before mtpro
\makeatletter
\@ifundefined{pdfoutput} %\% Definitely not using pdftex.
{ %\% Standard TeX
\usepackage[dvips]{graphicx,color}
}
{ %\% Running pdftex.
\ifnum\pdfoutput=0\relax% Are we outputting pdf?
\usepackage[dvips]{graphicx,color}
\fi
\ifnum\pdfoutput=1\relax% Are we outputting pdf?
\usepackage[pdftex]{graphicx,color}
\fi
}
\makeatother
\newcommand{\beq}{\begin{equation}}
\newcommand{\eeq}{\end{equation}}
\newcommand{\beqa}{\begin{eqnarray}}
\newcommand{\eeqa}{\end{eqnarray}}
\newcommand{\nn}{\nonumber \\}

\def \e {\mathrm{e}}

\def \la {\langle}
\def \ra {\rangle}
\def \s {\sigma}

\def \B {{\mathcal B}}

\def \I {{\mathbb{I}}}

\begin{document}
%%%%%%%%%%%%%%%%%%%%%%%%%%%%%%%%%%%%%%%%%%%%%%%%%%%%%%%%%%%%%%%%%%%%%%%%%%%%%%%
\title{ Topologically protected quantum gates for computation with
    non-Abelian anyons in the Pfaffian quantum Hall state}
%%%%%%%%%%%%%%%%%%%%%%%%%%%%%%%%%%%%%%%%%%%%%%%%%%%%%%%%%%%%%%%%%%%%%%%%%%%%%%%
%%%%%%%%%%%%%%%%%%%%%%%%%%%%%%%%%%%%%%%%%%%%%%%%%%%%%%%%%%%%%%%%%%%%%%%%%%%%%%%
%\title{Topologically protected Hadamard, phase and Controlled-NOT gates for
%quantum computation with non-Abelian anyons in the Pfaffian quantum Hall state}
%%%%%%%%%%%%%%%%%%%%%%%%%%%%%%%%%%%%%%%%%%%%%%%%%%%%%%%%%%%%%%%%%%%%%%%%%%%%%%%
%%%%%%%%%%%%%%%%%%%%%%%%%%%%%%%%%%%%%%%%%%%%%%%%%%%%%%%%%%%%%%%
%\title{A universal set of topologically protected gates for
%	quantum computation with Pfaffian qubits}
%%%%%%%%%%%%%%%%%%%%%%%%%%%%%%%%%%%%%%%%%%%%%%%%%%%%%%%%%%%%%%%

\author{Lachezar S. Georgiev}
\affiliation{Institute for Nuclear Research and Nuclear Energy,
	72 Tsarigradsko Chaussee, 1784 Sofia, Bulgaria}

\date{\today}
\begin{abstract}
We extend the topological quantum computation scheme using the  Pfaffian
quantum Hall state, which has been recently proposed  by Das Sarma et al.,
in a way that might potentially allow for the topologically protected
construction of a  universal set of quantum gates.
We construct, for the first time,  a topologically protected
Controlled-NOT gate which is entirely based on quasihole braidings
of Pfaffian qubits.
All single-qubit gates, except for the $\pi/8$ gate, are also explicitly
implemented by quasihole braidings.  Instead of the $\pi/8$ gate we try
to construct a topologically protected Toffoli gate, in terms of the
Controlled-phase gate and CNOT  or by a braid-group based
Controlled-Controlled-Z precursor.
We also give a topologically protected realization of the
Bravyi--Kitaev two-qubit gate $g_3$.
\end{abstract}


\maketitle
%\clearpage
%%%%%%%%%%%%%%%%%%%%%%%%%%%
\section{Introduction}
%%%%%%%%%%%%%%%%%%%%%%%%%%%
One of the spectacular features of the two-dimensional strongly correlated
electron systems is that they might possess quasiparticle  excitations
 obeying anyonic exchange statistics \cite{wilczek}.
Even more astonishing, the result of some quasiparticles exchanges
might not be just  phases but  non-trivial statistics matrices and such
quasiparticles are called  non-Abelian.
The most prominent non-Abelian candidate is the fractional quantum Hall (FQH)
state, which is now routinely observed at filling factor $\nu=5/2$ in ultra
high-mobility samples \cite{eisen2002,xia}.
This expectation is based on convincing analytical and numerical evidence
\cite{morf,rezayi-haldane,read-green,ivanov}
that this FQH state is most likely in the universality class of the Pfaffian
state constructed by Moore and Read \cite{mr} using correlation functions of
an appropriate $1+1$ dimensional conformal field theory (CFT).
While the experiments undoubtedly confirmed that FQH quasiparticle excitations
in general carry fractional electric charge \cite{picciotto}, the effects of
quantum statistics
are in principle much more difficult to observe. That is why even Abelian
fractional statistics has been experimentally tested only recently
\cite{goldman-stat}.
However, it turns out that the non-Abelian excitations might
be much easier to observe despite their more complicated structure.
Convincing proposals have been made for the detection of the
non-Abelian statistics of the
quasiparticles in the $\nu=5/2$ FQH state \cite{stern-halperin,bonderson-5-2,5-2AB}
and in the $\nu=12/5$ FQH state (anticipated to be the $k=3$ parafermion Hall state)
\cite{bonderson-12-5,chung-stone}.

In addition to its fundamental significance, the non-Abelian quantum statistics
might become of practical importance in another field of quantum theory, the
quantum computation \cite{nielsen-chuang}, which has been developing very
fast in
the recent years. Although the ideas behind quantum information processing
are simply based on the well established fundamental postulates of the
quantum theory, its
exponentially growing computational power could not have been used in
practice so far
due to the unavoidable obstacles caused by decoherence as a result
of interaction of the qubits with their environment.
Even the remarkable breakthrough based on quantum error correction algorithms
\cite{nielsen-chuang}
could not help creating a quantum computer with more than a few qubits.
Recently on this background  emerged the brilliant idea of
\textit{topological quantum computation}  (TQC)
\cite{kitaev-TQC,freedman-larsen-wang-TQC,sarma-freedman-nayak}.
Because the interactions
leading to noise and decoherence are presumably local we can try to
avoid them
by encoding quantum information non-locally, using some global
e.g., topological characteristics of the system. This started to be called
\textit{topological protection of qubit operations}---quantum information
is inaccessible to local interactions, because they
cannot distinguish between the computational basis states and hence cannot
lead to decoherence
\cite{kitaev-TQC,sarma-freedman-nayak,freedman-nayak-walker}.
That is why
topological gates are believed to be exact operations, which
might potentially allow to construct a truly scalable
 fault-tolerant quantum-computation platform.

The FQH liquid is a perfect candidate   for TQC
because it possesses a number of topological properties which are universal,
i.e., robust against the variations of the interactions details.
One could in principle use the braid matrices  representing the exchanges of
 FQH quasiparticles to implement arbitrary unitary transformations
\cite{kitaev-TQC,freedman-larsen-wang-TQC,kauffman-braid}.
However, because the single-qubit space is two-dimensional we need some
degeneracy
in order to implement the TQC scheme in terms of FQH quasiparticles,
i.e., we need degenerate spaces of quasiparticle correlation functions
with dimension at least 2. It is well known that the Abelian FQH quasiparticles
have degenerate spaces on non-trivial manifolds such as torus. Unfortunately,
these constructions are not appropriate for planar systems such  as the FQH
liquids though some diagonal two-qubit gates can be realized with abelian FQH
anyons as in Ref.~\onlinecite{averin-goldman}. On the other hand,
the non-Abelian FQH quasiparticles by definition
have degenerate spaces even in planar geometry and  are therefore better
suited for TQC.
Another virtue of the TQC scheme is its expected scalability---the solution of
the single
qubit operations problem might turn out to be the solution in general.
The only residual source of noise is due to thermally
activated quasiparticle--quasihole pairs which might execute
unwanted braids. Fortunately, these processes  are exponentially
suppressed at low temperature by the energy gap, which leads to
astronomical precision of quantum information encoding.

One significant step forward in the field of TQC has been done recently:
in a beautiful paper Das Sarma et al.
\cite{sarma-freedman-nayak,freedman-nayak-walker}
proposed to use the expected non-Abelian  statistics of the quasiparticles
in the Pfaffian FQH state
to construct an elementary qubit and execute a logical NOT gate on it
that could serve as a base for TQC. The  construction  of the NOT gate
is a fairly important issue in the field of quantum computation because it
underlies the qubit initialization procedure as well as the construction of
the single-qubit gates and the Controlled-NOT (CNOT) gate.

The FQH state at $\nu=5/2$ has one serious advantage compared to the other
candidates for TQC---it is the most stable state, i.e., the one with the
highest bulk energy gap, among all FQH states in which non-Abelian
quasiparticle statistics is expected to be realized.
On the other hand, one big disadvantage  of this state
is that the quasiparticles braiding matrices cannot be used alone for
universal quantum computation because the braid group representation
over the Ising model correlation functions
is finite \cite{read-JMP,sarma-freedman-nayak,TQC-NPB}. This has to be compared with
the FQH state observed at $\nu=12/5$ \cite{eisen2002,xia}, whose braid matrices
are expected to be universal, but whose energy gap is an order of magnitude
lower than that of the $\nu=5/2$ one.
The motivation for this paper is to demonstrate that
it might be possible to find a complete set of topologically
protected quantum gates in the Pfaffian state, though not all of them
would be realizable simply in terms of braidings. More precisely, we
explicitly construct by braiding  the single-qubit Hadamard gate $H$,
phase gate $S$, and the two-qubit CNOT gate, which generate a Clifford group
playing a central role in the error correction codes \cite{bravyi-kitaev-Clifford-QC}.
In addition we propose
a candidate for a topologically protected  three-qubit Toffoli gate
realized in terms of the two-qubit Controlled-$S$ gate and CNOT or by a
Controlled-Controlled-Z gate precursor realized by braiding.
This combination of gates is known to be sufficient for universal quantum
computation \cite{nielsen-chuang}.
%%%%%%%%%%%%%%%%%%%%%%%%%%%%%%%%%%%%%%%%%%%%%%%%%%%%%%%%%%%%%%%%%%%%%%%%%%%%%%
\section{One-qubit gates for Pfaffian qubits }
%%%%%%%%%%%%%%%%%%%%%%%%%%%%%%%%%%%%%%%%%%%%%%%%%%%%%%%%%%%%%%%%%%%%%%%%%%%%%%
The Pfaffian qubit is constructed in terms of the wave functions
for the excitations containing 4 quasiholes
\cite{sarma-freedman-nayak,freedman-nayak-walker}.
For fixed positions of the quasiholes there are two independent functions
forming the computational basis,
which can be conveniently written as \cite{nayak-wilczek}
\beq\label{01}
|0\ra,|1\ra \Leftrightarrow \Psi_{4\mathrm{qh}}^{(0,1)}=
\frac{\left(\eta_{13}\eta_{24}\right)^{\frac{1}{4}}}{\sqrt{1\pm \sqrt{x}}}
\left(\Psi_{(13)(24)}\pm \sqrt{x}\,  \Psi_{(14)(23)} \right)
\eeq
where $\eta_1, \ldots, \eta_4$ are the quasiholes positions,
$x$ is a CFT invariant  crossratio \cite{CFT-book}
\beq \label{x}
x=\frac{\eta_{14}\eta_{23}}{\eta_{13}\eta_{24}}, \quad
\mathrm{and} \quad \eta_{ab}=\eta_a-\eta_b.
\eeq
The explicit form of the  functions $\Psi_{(ab)(cd)}$, which are computed as
Pfaffians, will not be needed here, we shall only use that they are
 single-valued in the positions of the quasiholes \cite{nayak-wilczek,5-2}.
  A convenient way to fix the positions
of the quasiholes is to localize them on antidots \cite{sarma-freedman-nayak}.
The readout of the qubit state could be efficiently done  by measuring
the interference pattern of the
longitudinal conductance in an electronic Mach--Zehnder interferometer
\cite{sarma-freedman-nayak,freedman-nayak-walker,Mach-Zehnder-Nature}.

The basic idea of the TQC scheme of Ref.~\onlinecite{sarma-freedman-nayak}
is that the single-qubit gates can be realized in
terms of the two-dimensional exchange matrices of 4 quasiholes,
defined here in the basis (\ref{01}).
In order to obtain the exchange matrix $R_{a, \, a+1}^{(4)}$ we first exchange
the coordinates $\eta_a \leftrightarrow \eta_{a+1}$ in counter-clockwise
direction,
which changes the crossratio (\ref{x}) in a tractable way \cite{TQC-NPB},
and then use the
analytic properties of the wave functions (\ref{01}) to extract
$R_{a, \ a+1}^{(4)}$.
Thus we find  the elementary braid  matrices \cite{TQC-NPB} in the basis
(\ref{01})
\beq \label{R-4}
R_{12}^{(4)}=R_{34}^{(4)}=\left( \begin{matrix}1 & 0 \cr 0 & i \end{matrix}\right),
\quad
R_{23}^{(4)}=\frac{\e^{i\frac{\pi}{4}} }{\sqrt{2}} \left(
\begin{matrix}    1 & -i \cr -i &   1\end{matrix}\right) ,
\eeq
which together with their inverses, generate the entire representation of the
braid group $\B_4$. The generators $B_i$,  $i=1,\ldots, n-1$, of the
braid group
$\B_n$  satisfy in general the Artin relations \cite{birman}
\beqa \label{artin}
	B_i B_j &=&  B_j B_i, \qquad \qquad \mathrm{for}  \quad |i-j|\geq 2 \nn
  B_i B_{i+1} B_i &=&  B_{i+1} B_i B_{i+1}, \quad \mathrm{with}
	\quad B_i=R_{i,i+1} \in \B_n.
\eeqa
In our case of planar geometry  the exchange matrices
$B_i=R_{i,i+1}^{(n)}$   should satisfy  one more relation \cite{birman}
\[
B_1 B_2 \cdots B_{n-2} B_{n-1}^2 B_{n-2} \cdots  B_2 B_1 = \I ,
\]
as is appropriate for the representation of the braid group on the sphere.
Because this is satisfied in the Pfaffian TQC scheme  only up to a phase
factor,  the representations of the braid groups are generically
 projective \cite{birman}.

\begin{figure}[htb]
\centering
\includegraphics*[bb=10 330 595 520,width=8cm]{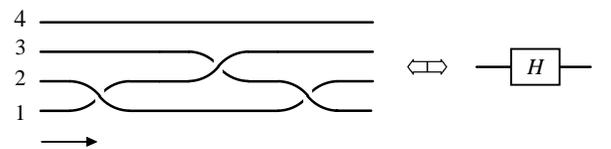}
 \caption{Braiding diagram for the Hadamard gate (\ref{H}) and its symbol
 	(on the right)  in standard	quantum-computation notation.}
\label{fig:H}
\end{figure}

\begin{figure}[htb]
\centering
\includegraphics*[bb=0 330 595 530,width=7.5cm]{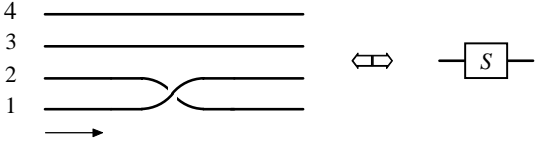}
 \caption{Braiding diagram for the phase  gate  $S$}
\label{fig:S}
\end{figure}

In other words, everything which can be obtained as a result of
quasiparticle exchanges could be expressed in terms of the elementary exchange
matrices (\ref{R-4}).
For example, skipping throughout overall phases, the Hadamard gate
\cite{nielsen-chuang} can be executed by 3 elementary braids (skipping
the superscript ``(4)'' of $R$)
\beq \label{H}
H\simeq R_{13}R_{12}^2 = R_{12}^{-1} R_{23} R_{12}^{-1}  \simeq
\frac{1}{\sqrt{2}}
\left( \begin{matrix}1 & \ \ \  1 \cr 1 & -1 \end{matrix}\right)
\eeq
and its braid diagram in terms of the elementary exchanges is shown on
Fig.~\ref{fig:H}.
%%%%%%%%%%%%%%%%%%%%
%%%%%%%%%%%%%%%%%%%
Similarly, the Pauli $X$ gate, first implemented in
Ref.~\onlinecite{sarma-freedman-nayak},
and the phase gate \cite{nielsen-chuang} $S$  are expressed as
\[
X = R_{23}^2=
\left( \begin{matrix}0 & 1 \cr 1 & 0\end{matrix}\right), \quad
S = R_{12}= R_{34} =
\left( \begin{matrix}1 & 0 \cr 0 & i\end{matrix}\right)
 \]
%%%%%%%%%%%%%%%%%%%%
\begin{figure}[htb]
\centering
\includegraphics*[bb=10 340 595 520,width=7.5cm]{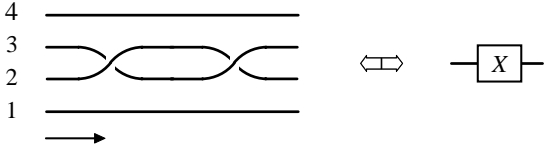}
 \caption{Braiding diagram for the Pauli $X$ gate.}
\label{fig:X}
\end{figure}
%%%%%%%%%%%%%%%%%%%
and their braid diagrams are shown on Figs.~\ref{fig:X} and \ref{fig:S}, respectively.
%%%%%%%%%%%%%%%%%%%%
%%%%%%%%%%%%%%%%%%%
Notice that $S^2=Z$, where $Z$ is Pauli $Z$ gate, as it should be \cite{nielsen-chuang}.
The last remaining single-qubit gate, the $\pi/8$ gate $T=\mathrm{diag}(1,\e^{i\pi/4})$,
which is necessary
for approximating arbitrary single-qubit gate,	cannot be realized by exact
braidings because $\det T =\e^{i\pi/4}$, while $\det R_{a,\ a+1}^{(4)}=i$.
Instead, we will try  to construct later the Toffoli gate
\cite{nielsen-chuang}.
%%%%%%%%%%%%%%%%%%%%%%%%%%%%%%%%%%%%%%%%%%%%%%%%%%%%%%%%%%%%%%%%%%%%%%%%%%%%%%
\section{Two-qubit gates: Controlled-NOT}
\begin{figure}[htb]
\centering
\includegraphics*[bb=90 570 510 660,width=9cm]{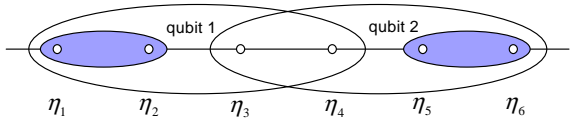}
 \caption{Two qubits, realized by quasiholes with coordinates
$(\eta_1,\eta_2,\eta_3,\eta_4)$ and $(\eta_3,\eta_4,\eta_5,\eta_6)$.
The state of qubit 1 is determined by the quasiholes with positions
$(\eta_1,\eta_2)$, while the state of qubit 2 by $(\eta_5,\eta_6)$
and that is why these two groups are shaded.}
\label{fig:2qubits}
\end{figure}

%%%%%%%%%%%%%%%%%%%%%%%%%%%%%%%%%%%%%%%%%%%%%%%%%%%%%%%%%%%%%%%%%%%%%%%%%%%%%%
It is natural to use 6 quasiholes, as shown on Fig.~\ref{fig:2qubits}, to
construct two-qubit gates, because the 6-quasiholes wave functions
form a 4-dimensional space
(in general the dimension of the space of wave functions with $2n$ quasiholes
with fixed positions \cite{nayak-wilczek} is $2^{n-1}$).
For the selected class of the controlled two-qubit operations
\cite{nielsen-chuang} the first two
quasiholes determine the state of the control qubit  while the last two
form the target qubit.
%%%%%%%%%%%%%%%%%%%%
%%%%%%%%%%%%%%%%%%%
Just like with 4 quasiholes the state of a qubit is
$|0\ra \leftrightarrow \s_+\s_+$ if the two
quasiholes fuse to the unit operator, or $|1\ra\leftrightarrow \s_+\s_-$
if they fuse to form a Majorana fermion \cite{sarma-freedman-nayak}.
Here we use that the neutral part of the quasihole is represented by
the chiral spin field $\s_{\pm}$ of the Ising model, with CFT dimension
$1/16$, and the subscript $\pm$ denotes its fermion parity \cite{5-2,TQC-NPB}.
Thus we may define the two-qubit basis  by
\beqa \label{2qb-basis}
&&|00\ra \equiv \la \s_+ \s_+\s_+\s_+\s_+\s_+\ra, \ \
|01\ra \equiv \la \s_+ \s_+\s_+\s_-\s_+\s_-\ra \nn
&&|10\ra \equiv \la \s_+ \s_-\s_+\s_-\s_+\s_+\ra,  \ \
|11\ra \equiv \la \s_+ \s_-\s_+\s_+\s_+\s_-\ra  . \quad \
\eeqa
which is convenient since if we fuse the first two quasiholes this would
project to the second qubit, while if we fuse the last two quasiholes
we project to the first qubit \cite{TQC-NPB}, i.e.,
\beq \label{project}
	|\alpha\beta\ra \mathop{\to}_{\eta_1\to\eta_2} |\beta\ra , \quad
	|\alpha\beta\ra \mathop{\to}_{\eta_5\to\eta_6} |\alpha\ra  .
\eeq
Then the state of the third and the fourth quasiholes are fixed by the
conservation
of the fermion parity, i.e., if $e_i$ is the parity of $\s_{e_i}$  then
$e_3e_4=e_1e_2e_5e_6$ \cite{TQC-NPB}.
Thus we have only 4 independent states in the space of 6-quasiholes
with fixed positions, which correspond to Eq.~(\ref{2qb-basis}).

Using the two-qubit states  definition  (\ref{2qb-basis})
and the projection (\ref{project}) it is easy
to find the exchange matrices for 6 quasiholes from those for 4 quasiholes.
For example, to obtain  $R_{12}$, we may first fuse the last two quasiholes
$\eta_5 \to \eta_6$ and then identify this braid matrix with the tensor product
$R^{(6)}_{12}=R^{(4)}_{12}\otimes \I_2$, where $\I_2$ is the two-dimensional
unit matrix and the superscript of $R$ shows the number of quasiholes.
Thus the elementary  6-quasiholes braid matrices are
\cite{TQC-NPB}
\beqa \label{R-6-1}
&&R_{12}^{(6)} = \left(
\begin{matrix}
1 & 0 & 0 & 0 \cr  0 & 1 & 0 & 0 \cr 0 & 0 & i & 0 \cr 0 & 0 & 0 & i
\end{matrix} \right),
\
R_{23}^{(6)}  =\frac{\e^{i\frac{\pi}{4}}}{\sqrt{2}}\left(
\begin{matrix}1 & 0 & -i & 0 \cr  0 & 1 & 0 & -i \cr -i & 0 & 1 & 0 \cr 0 & -i & 0 & 1
\end{matrix} \right) \quad  \\      \label{R-6-2}
&&R_{34}^{(6)}  = \left(
\begin{matrix}1 & 0 & 0 & 0 \cr  0 & i & 0 & 0 \cr 0 & 0 & i & 0 \cr 0 & 0 & 0 & 1
\end{matrix} \right),
\
R_{45}^{(6)}  =\frac{\e^{i\frac{\pi}{4}}}{\sqrt{2}}\left(
\begin{matrix}1 & -i  & 0 & 0 \cr  -i & 1 & 0 & 0 \cr 0 & 0 & 1 & -i \cr 0 & 0 & -i & 1\end{matrix} \right)
\\    \label{R-6-3}
&&R_{56}^{(6)}  = \left(
\begin{matrix}1 & 0 & 0 & 0 \cr  0 & i & 0 & 0 \cr 0 & 0 & 1 & 0 \cr 0 & 0 & 0 & i
\end{matrix} \right).
\eeqa
It is not difficult to check that the exchange matrices (\ref{R-6-1}),
(\ref{R-6-2}) and (\ref{R-6-3}) satisfy the Artin relations
(\ref{artin}) for the braid group $\B_6$.

\begin{figure}[htb]
\centering
\includegraphics*[bb=50 345 578 490,width=8.7cm]{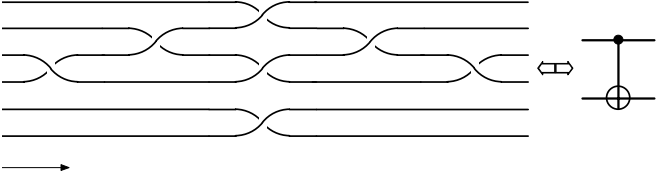}
 \caption{The braid diagram for the  Controlled-NOT gate executed by
7 elementary 6-quasiparticle braids corresponding to Eq.~(\ref{CNOT}).}
\label{fig:CNOT-braid}
\end{figure}

One of the main advantages of our two-qubit construction (\ref{2qb-basis}),
shown on Fig.~\ref{fig:2qubits},
is that it is straight forward to execute various two-qubit gates by braiding.
For example,
using the explicit braid matrices from Eqs.~(\ref{R-6-1}), (\ref{R-6-2}) and (\ref{R-6-3}),
it is  easily verified that the CNOT gate can be implemented, in the basis
(\ref{2qb-basis}), by
\beq \label{CNOT}
 \mathrm{CNOT} = R_{34}^{-1} R_{45} R_{34}R_{12}R_{56} R_{45}  R_{34}^{-1}
 =\left(
\begin{matrix}1 & 0 & 0 & 0 \cr  0 & 1 & 0 & 0 \cr 0 & 0 & 0 & 1 \cr 0 & 0 & 1 & 0
\end{matrix} \right)   ,
\eeq
whose braid  diagram is shown   on Fig.~\ref{fig:CNOT-braid}.
%%%%%%%%%%%%%%%%%%%%%
%%%%%%%%%%%%%%%%%%%
An equivalent realization would be
\[
\mathrm{CNOT}=R_{56} R_{45}  R_{56}^{-1} R_{34}^{-1}  R_{12} R_{45} R_{56}
\]
giving precisely the same result as in Eq.~(\ref{CNOT}).
It is quite remarkable that the CNOT gate can be implemented entirely in
terms of 6-quasiholes braidings which certainly guarantees its exactness and
topological protection.
The best achievement in this direction so far has been the proposed realization
\cite{freedman-nayak-walker} of a Controlled-Z gate\footnote{the CNOT gate can
be obtained from CZ with the help of the Hadamard gate acting on the target
qubit \cite{nielsen-chuang}}
precursor
by taking $(\eta_5,\eta_6)$ around $\eta_1$, which corresponds in our approach
to $R_{15}^2R_{16}^2\simeq R_{56}^2$, i.e., expressed in terms of elementary
braids (skipping the superscript ``$(6)$'' of $R$)
\beqa
\widetilde{\mathrm{CZ}} & = &
R_{12}^{-1}R_{23}^{-1} R_{34}^{-1} R_{45}R_{56}^2 R_{45}R_{34}R_{23}R_{12} \nn
 & = &  \mathrm{diag}(1,-1,1,-1).   \nonumber
\eeqa
However,  this must be supplemented by the Bravyi--Kitaev procedure
\cite{freedman-nayak-walker} in which the quasiholes
with positions $\eta_1$ and $\eta_2$ are split only if their qubit
is in the state $|1\ra$, thus removing the minus sign on the second row.

While not necessary for our TQC scheme,  we give below for reference
another important two-qubit gate, the $g_3$ gate of Bravyi--Kitaev
\cite{freedman-nayak-walker}, in terms of the 6-quasihole exchange matrix
(skipping again the superscript ``$(6)$'' of $R$)
\beqa
R_{16} &=& R_{12}^{-1} R_{23}^{-1} R_{34}^{-1} R_{45}^{-1}
	 R_{56} R_{45}R_{34}R_{23}R_{12}  \nn
&\simeq&\frac{1}{\sqrt{2}}\left(
\begin{matrix}1 & 0 & 0 & -i \cr 0 & 1 & -i  & 0 \cr 0 & -i & 1 & 0 \cr
	-i & 0 & 0 &  1 \end{matrix} \right) = g_3 .
\eeqa
%%%%%%%%%%%%%%%%%%%%%%%%%%%%%%%%%%%%%%%%%%%%%%%%%%%%%%%%%%%%%%%%%%%%%%%%%%%%%%
\section{Three-qubit gates}
%%%%%%%%%%%%%%%%%%%%%%%%%%%%%%%%%%%%%%%%%%%%%%%%%%%%%%%%%%%%%%%%%%%%%%%%%%%%%%
In order to make our scheme universal we may try, instead of using
 $T$,   to construct the Toffoli gate
Controlled-Controlled-NOT (CCNOT) in terms of the Controlled-Controlled-Z gate
(CCZ) and the Hadamard gate acting on the target qubit
\beq \label{CCNOT-CCZ}
\mathrm{CCNOT}=  H_3 \  \mathrm{CCZ} \   H_3, \quad \mathrm{with} \quad
H_3= \I_2 \otimes\I_2 \otimes H,
\eeq
where $\mathrm{CCZ}=\mathrm{diag}(1,1,1,1,1,1,1,-1)$.
When the third qubit is defined by 2 more quasiholes at $\eta_7$ and
$\eta_8$  we can express (using the fusion rules of the non-Abelian quasiholes)
the exchange matrices for 8 quasiholes recursively in terms of those
for 6 quasiholes as follows \cite{TQC-NPB}:
\beqa \label{R-8-1}
R_{12}^{(8)}&=&R_{12}^{(6)}\otimes \I_2, \quad
R_{23}^{(8)}=R_{23}^{(6)}\otimes \I_2, \nn
R_{34}^{(8)}&=&\mathrm{diag}(1,  i, i, 1, i, 1, 1, i), \nn
R_{45}^{(8)}&=&R_{45}^{(6)}\otimes \I_2, \quad
R_{56}^{(8)}= R_{56}^{(6)} \otimes\I_2
\eeqa
\beq \label{R-8-2}
R_{67}^{(8)}=\frac{\e^{i\frac{\pi}{4}}}{\sqrt{2}}\left(
\begin{matrix}1 & 0 & 0 & -i & 0 & 0 & 0 & 0 \cr 0 & 1 & -i & 0 & 0 & 0  & 0 & 0 \cr
0 & -i & 1 & 0 & 0 & 0  & 0 & 0 \cr -i & 0 & 0 & 1 & 0 & 0  & 0 & 0 \cr
0 & 0 & 0 & 0 & 1 & 0 & 0 & -i \cr 0 & 0 & 0 & 0 & 0 & 1  & -i & 0 \cr
0 & 0 & 0 & 0 & 0 & -i &  1 & 0 \cr 0 & 0 & 0 & 0 & -i & 0 & 0  & 1
\end{matrix}
\right),
\eeq
and
\beq \label{R-8-3}
R_{78}^{(8)}=\I_2 \otimes R_{56}^{(6)} .
\eeq
The direct check shows that the exchange matrices
(\ref{R-8-1}), (\ref{R-8-2}) and (\ref{R-8-3}) satisfy the Artin
relations (\ref{artin}) for the braid group $\B_8$.

Now we can explicitly construct the three-qubit Hadamard gates in terms
of the exchange matrices  for 8 quasiholes (skipping the superscript
``$(8)$'' of $R$), e.g., the Hadamard gate acting on the first qubit is
\[
H_1=H\otimes \I_2\otimes \I_2 \simeq R_{12}^{-1}R_{23}^{-1}R_{12}^{-1}
\]
and that acting on the second qubit is
\[
H_2=\I_2\otimes H\otimes \I_2 \simeq R_{56}^{-1}R_{45}^{-1}R_{56}^{-1} .
\]
Instead of the Hadamard  gate $H_3$ acting on the third qubit, which is more
difficult to construct, we can use in Eq.~(\ref{CCNOT-CCZ})
\beq
\widetilde{H}_3 = R_{78}R_{45}R_{56}R_{67}^{-1}R_{56} R_{45}R_{78} \simeq
\I_2\otimes \I_2 \otimes H,
\eeq
that still does the job of expressing  CCNOT in terms of CCZ, while
differing slightly from $H_3$.
Note that, due to a specific feature built into  the braid group
representation,
it may not be always possible to represent exactly the single- and
two- qubit gates in the three-qubit basis \cite{TQC-NPB}.

\begin{figure}[htb]
\centering
\includegraphics*[bb=70 560 590 690,width=7.5cm]{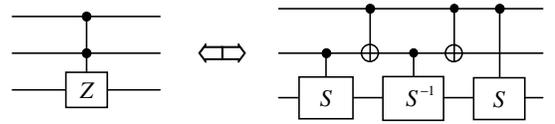}
 \caption{The Controlled-Controlled-Z gate realized by the Controlled-S gate and CNOT
 	(see Ref.~\onlinecite{nielsen-chuang} for the notation).}
\label{fig:CCZ}
\end{figure}

\begin{figure}[htb]
\centering
\includegraphics*[bb=0 500 570 650,width=7.5cm]{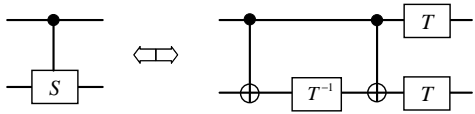}
 \caption{The Controlled-$S$ gate in terms of the $\pi/8$ gate $T$ and CNOT.}
\label{fig:CS}
\end{figure}

Next, because $Z=S^2$, we could construct \cite{nielsen-chuang} the
CCZ gate with the circuit shown on Fig.~\ref{fig:CCZ},
%%%%%%%%%%%%%%%%%%%%
%%%%%%%%%%%%%%%%%%%
using the CNOT gate (\ref{CNOT})
 and  the Controlled-S gate  (CS) defined in the basis (\ref{2qb-basis}) by
$\mathrm{CS}=\mathrm{diag}(1,1,1,i)$.
Obviously CS cannot be expressed directly as a product of 6-quasiholes
exchange matrices because $\det \mathrm{CS} =i$, while
$\det \left( R_{a, \ a+1}^{(6)}\right)=-1$.
We can  try to realize the CS gate using as a braid-group based
precursor the braid matrix $R_{56}^{(6)}$ in Eq.~(\ref{R-6-3}) supplementing
it by
the Bravyi--Kitaev construction to split the first qubit into two
separated
quasiholes at $\eta_1$ and $\eta_2$ only if it is in the state $|1\ra$.
In the standard quantum computation context, the CS gate can be realized
with the help of the $\pi/8$  gate $T$ and CNOT, as shown on
Fig.~\ref{fig:CS}, which demonstrates that
the use of the Toffoli gate is equivalent to the use of $T$.
%%%%%%%%%%%%%%%%%%%%
%%%%%%%%%%%%%%%%%%%

The construction of the CNOT gate in Eq.~(\ref{CNOT}) suggests another
possibility to realize the CCZ gate using as a braid-group based precursor the
diagonal matrix
\beqa \label{CCZ-pre}
\widetilde{\mathrm{CCZ}} &\equiv &
R_{12}^{(8)}  R_{34}^{(8)} R_{56}^{(8)} R_{78}^{(8)} \nn
&\simeq& \mathrm{diag}\left(-1,1,1,1,1,1,1,-1\right)
\eeqa
and apply again the Bravyi--Kitaev construction
that the first qubit is split into two charge-$1/4$ quasiholes only if it
is in the state $|1\ra$ in order to remove the minus sign on the first row.
This is precisely the same situation as with the CZ precursor of
Ref.~\onlinecite{freedman-nayak-walker}, where the Bravyi--Kitaev construction
removing the minus sign on the second row of their gate $g_2$,  is
believed to be realizable by tilted interferometry in planar geometry.
Note that the tilted interferometry realization of the Bravyi--Kitaev
construction \cite{freedman-nayak-walker} for Eq.~(\ref{CCZ-pre}) could
eventually provide us with a topologically protected CCZ gate, hence, with a
universal set of topologically protected gates that are sufficient for
universal quantum computation with Pfaffian qubits.

Alternatively,  to make our scheme universal, we could use the unprotected
$\pi/8$ gate $U_P$ of Ref.~\onlinecite{freedman-nayak-walker}, or use the magic
states   of Ref.~\onlinecite{bravyi-kitaev-Clifford-QC}, such as
$|H\ra = \cos\left(\pi/8\right)|0\ra + \sin \left(\pi/8\right)|1\ra$,
which in principle allow to build  the $T$ gate, while being simpler to
construct than the $\pi/8$ gate itself.
%%%%%%%%%%%%%%%%%%%%%%%%%%%%%%%%%%%%%%%%%%%%%%%%%%%%%%%%%%%%%%%%%%%%%%%%%%%%%%
\section{Conclusion}
%%%%%%%%%%%%%%%%%%%%%%%%%%%%%%%%%%%%%%%%%%%%%%%%%%%%%%%%%%%%%%%%%%%%%%%%%%%%%%
We demonstrated that it might indeed be possible to build
a universal quantum computer with topologically protected quantum gates
realized by quasiparticle braidings in the FQH state at $\nu=5/2$. Our quantum
computation scheme is based on the Hadamard gate $H$, the phase gate $S$,
the CNOT and possibly the Toffoli gate expressed in terms the Controlled-$S$
gate. If in addition the CCZ gate could be realized by tilted interferometry
from the braid-group based precursor proposed here, then we would be able
to construct the Toffoli gate and hence implement arbitrary quantum gates
in a topologically protected way.

We should stress that it is not completely trivial to embed all one-qubit and
two-qubit gates, realized in this paper,  into systems with three or
more qubits because the construction of some of those gates is based on
specific properties of
the generators of braid groups $\B_4$ and $\B_6$, which are not directly
generalized for $\B_n$ with $n\geq 8$. Nevertheless, the embedding of the Clifford gates
$H$, $S$ and CNOT into a three-qubit system seems  possible though that
might require some more work.

%%%%%%%%%%%%%%%%%%%%%%%%%%%%
I thank Chetan Nayak  and Michael Geller for useful discussions.
This work has been partially supported by the FP5-EUCLID Network Program
 HPRN-CT-2002-00325 and by the BG-NCSR under Contract No. F-1406.
%%%%%%%%%%%%%%%%%%%%%%%%%%
%\bibliography{Z_k,/home/lgeorg/documents/publications/my}
\bibliography{my,TQC,FQHE,Z_k}
\end{document}